\documentclass[conference]{IEEEtran}

\IEEEoverridecommandlockouts

\usepackage{cite}
\usepackage{amsmath,amssymb,amsfonts}
\usepackage{algpseudocode} %
\usepackage{algorithm}     %
\usepackage{graphicx}
\usepackage{textcomp}
\usepackage{color,soul}
\usepackage{lipsum} %
\usepackage{xurl}

\usepackage{booktabs} %
\usepackage{array} %
\usepackage{makecell} %
\usepackage{url}
\usepackage{booktabs}
\usepackage{amssymb}
\usepackage{tabularx}
\usepackage{optidef}
\usepackage{graphicx}
\usepackage{wrapfig}
\usepackage{afterpage}
\usepackage{float}
\usepackage{multirow}
\usepackage{array}
\usepackage{hhline}
\usepackage{arydshln}
\usepackage{tabularx}
\usepackage{makecell}
\usepackage{subcaption}
\usepackage{booktabs}
\usepackage{indentfirst}
\usepackage{adjustbox}
\usepackage{amsmath}
\usepackage{caption}
\usepackage{soul}
\usepackage{soulpos}
\DeclareMathOperator*{\argmax}{arg\,max}
\DeclareMathOperator*{\argmin}{arg\,min}

\usepackage{bm} %
\usepackage{mathptmx} %

\usepackage{ifthen} %
\usepackage{xcolor} %
\usepackage{ulem}   %

\newcommand{\showchanges}{0}

\newcommand{\added}[1]{\ifthenelse{\equal{\showchanges}{1}}{\textcolor{blue}{#1}}{#1}}
\newcommand{\deleted}[1]{\ifthenelse{\equal{\showchanges}{1}}{\textcolor{red}{\sout{#1}}}{}}

\usepackage{bm}
\begin{document}

\title{%
Energy-Efficient Deep Learning for Traffic Classification on Microcontrollers
\thanks{This work was partially supported by project SERICS (PE00000014) under the MUR National Recovery and Resilience Plan funded by the European Union - NextGenerationEU.}
}

\author{
    \IEEEauthorblockN{Adel Chehade, Edoardo Ragusa, Paolo Gastaldo, and Rodolfo Zunino}
    \IEEEauthorblockA{
        DITEN, University of Genoa, Italy \\
        Email: adel.chehade@edu.unige.it, \{edoardo.ragusa, paolo.gastaldo, rodolfo.zunino\}@unige.it
    }
}

\maketitle

\begin{abstract}
In this paper, we present a practical deep learning (DL) approach for energy-efficient 
traffic classification (TC) on resource-limited microcontrollers, 
which are widely used in IoT-based smart systems and communication networks. Our objective is to balance accuracy, computational efficiency, and real-world deployability. To that end, we develop a lightweight 1D-CNN, optimized via hardware-aware neural architecture search (HW-NAS), which achieves 96.59\% accuracy on the ISCX VPN–NonVPN dataset with only 88.26K parameters, a 20.12K maximum tensor size, and 10.08M floating-point operations (FLOPs). Moreover, it generalizes across various TC tasks, with accuracies ranging from 94\% to 99\%. To enable deployment, the model is quantized to INT8, suffering only a marginal 1–2\% accuracy drop relative to its Float32 counterpart. We evaluate real-world inference performance on two microcontrollers: the high-performance STM32F746G-DISCO and the cost-sensitive Nucleo-F401RE. The deployed model achieves inference latencies of 31.43~ms and 115.40~ms, with energy consumption of 7.86~mJ and 29.10~mJ per inference, respectively. These results demonstrate the feasibility of on-device encrypted traffic analysis, paving the way for scalable, low-power IoT security solutions.
\end{abstract}


\section{Introduction}
\label{sec:introduction}

Traffic classification (TC) plays a critical role in ensuring the security and efficiency of modern networks \cite{nunez2024novel_modern_networks}. It enables key tasks such as intrusion detection, network optimization, and resource allocation \cite{ports_dai2023glads}. Yet, the rise of encryption has rendered traditional methods like Deep Packet Inspection (DPI) and port-based classification ineffective, as they rely on payload visibility or static port mappings \cite{seydali2023cbs, survey_DONG2024128444}. This challenge is further compounded by the reliance on cloud-based solutions, which require transferring raw traffic data to centralized servers for analysis \cite{shahin2024advancing_iot_dataflow}. Such cloud-based approaches introduce significant drawbacks, including increased latency, high bandwidth consumption, and privacy concerns. Meanwhile, microcontrollers—widely used in IoT and embedded systems—are becoming integral to connected infrastructures, driving demand for efficient on-device processing.

This work addresses these challenges by targeting microcontroller-based IoT devices as platforms for performing local TC. These systems offer real-time decision-making while minimizing network overhead and privacy risks \cite{shahin2024advancing_iot_dataflow}. However, microcontrollers are resource-constrained, with limitations on memory, computational power, and energy efficiency. These factors necessitate tailored optimizations to ensure effective classification within strict hardware limits.

Accordingly, network TC can occur at packet, flow, or session granularity \cite{survey_wang2022machine}. Packets, as the smallest transmission units, carry limited context when analyzed alone. Flows group packets unidirectionally for broader context but may still miss complete endpoint interactions. Session-level classification, however, captures bidirectional exchanges, making it ideal for encrypted traffic scenarios by leveraging temporal and contextual information from network activities \cite{zou2024novel_residual,wang2017end}.
Since encryption limits access to packet payloads, many approaches rely on handcrafted features \cite{survey_DONG2024128444, survey_wang2022machine}. However, feature engineering adds preprocessing delays and requires domain expertise, complicating real-time deployment. Directly processing raw traffic data allows models to learn representations from encrypted communication without predefined features.

Based on this, deep neural networks (DNNs) have proven effective for TC by automating feature extraction and effectively analyzing complex network patterns \cite{lotfollahi2020deep}. However, DNNs are computationally intensive and memory-demanding, making their deployment on microcontrollers a challenge. Neural Architecture Search (NAS) has emerged as a promising framework for automating the design of neural networks across various deep learning (DL) domains. Although effective in optimizing performance metrics such as accuracy, traditional NAS ignores hardware constraints imposed by IoT platforms.

To overcome this limitation, we use Hardware-Aware Neural Architecture Search (HW-NAS), 
which extends NAS by explicitly modeling the memory, computational, and energy constraints inherent to microcontrollers \cite{1000papersNAS}\cite{chitty2023neural_ieee-access}. 
This approach yields lightweight yet accurate DNNs that meet strict resource limits. The model, optimized using the ISCX VPN-nonVPN dataset \cite{iscxvpnnonvpn}, is deployed and evaluated on microcontrollers, including the Nucleo F401RE and STM32F746G-DISCO, focusing on real-world performance. Moreover, it maintains strong accuracy across multiple TC tasks, confirming its adaptability for broader network analysis.

The key contributions of this paper are as follows:

\begin{itemize}
    \item We automatically design and optimize a compact 1D-CNN using HW-NAS for session-level TC, ensuring a balance between accuracy and efficiency while remaining lightweight enough for deployment on microcontrollers. The final model operates with only 88.26K parameters, a maximum tensor size of 20.12K, and 10.08M floating point operations (FLOPs), which makes it computationally efficient for embedded systems.
    
    \item We demonstrate real-world feasibility by deploying the optimized model on two embedded platforms: the STM32F746G-DISCO, a high-performance board, and the Nucleo-F401RE, a cost-sensitive alternative. Experimental results show that the model runs efficiently on both devices, achieving inference latencies of 31.43 ms and 115.40 ms, with energy consumption of 7.86 mJ and 29.10 mJ per inference, respectively. These results confirm the model's capability to operate within strict resource constraints while meeting the real-time demands of embedded network applications.
    
    \item The model, optimized using the ISCX VPN–NonVPN dataset, achieves 96.59\% accuracy on the primary task and generalizes well across multiple TC challenges, including VPN protocol encapsulation detection (99.95\%), VPN traffic type classification (99.19\%), non-VPN traffic type classification (94.17\%), and network usage categorization (96.94\%). This demonstrates that high-performance TC remains feasible even under the limitations of microcontroller-based platforms.
\end{itemize}

\section{Related Work}

\subsection{Conventional Traffic Classification Techniques}

Early TC methods relied on port-based classification and Deep Packet Inspection (DPI). Port-based classification maps network traffic to applications using fixed port numbers. However, modern networks, with their use of dynamic and private ports, port obfuscation, and HTTP tunneling, have rendered this approach largely ineffective \added{as noted in} \cite{survey_DONG2024128444}. DPI, which examines packet payloads for patterns or signatures, was effective for unencrypted traffic but is unsuitable for encrypted communication, where payloads are inaccessible. Additionally, DPI is resource-intensive and raises privacy concerns, limiting its feasibility in contemporary environments \cite{dpi_hongke2022dpi, survey_DONG2024128444}.

\subsection{Machine Learning Techniques}

Machine learning (ML) has provided an alternative to conventional methods by leveraging statistical patterns in network traffic to classify encrypted communication without inspecting packet payloads. Early ML-based approaches used models such as Naïve Bayes, Random Forests, and k-NN to classify traffic based on handcrafted features like packet size and flow duration\added{, as in}~\cite{ml_skype_dong2019retracted, ml_rf_supervised_zhai2018random, iscxvpnnonvpn}. While these methods showed reasonable accuracy, they required significant manual effort for feature selection and engineering to adapt to evolving network traffic patterns, limiting their scalability and robustness.

DL has advanced TC by automating feature extraction, enabling the analysis of raw network traffic without manual feature engineering. Techniques such as convolutional neural networks (CNNs) and recurrent neural networks (RNNs) have demonstrated high accuracy in encrypted traffic classification (ETC) \cite{survey_DONG2024128444}. For example, 1D-CNNs have been effectively applied in studies such as \cite{lotfollahi2020deep, wang2017end}, achieving competitive accuracy in analyzing raw traffic data. Additionally, \added{the work in}~\cite{lu2021iclstm} used Inception and Long Short-Term Memory (LSTM) networks to capture spatial and temporal features, further improving classification accuracy. However, the computational demands of these models remain a barrier to their deployment in resource-constrained environments like IoT devices.

The CBS model, \added{from the study in}~\cite{seydali2023cbs}, integrated CNNs, attention-based Bi-LSTMs, and stacked autoencoders to leverage spatial, temporal, and statistical information for ETC. While this approach achieved high accuracy, its computational cost made it unsuitable for real-time applications. Capsule Networks (CapsNet) were explored \added{by the authors in}~\cite{cui2019session}, prioritizing accuracy over efficiency, making them difficult to deploy on constrained platforms. Attention-based models, \added{such as the work in}~\cite{attention_spatio_hu2023network}, also improve representation but add significant overhead, limiting their practicality for IoT devices.

\subsection{Neural Architecture Search and Hardware-Aware NAS}  
NAS has gained attention as an effective framework for automating DNN design. It explores candidate architectures to optimize performance metrics \cite{zoph2016neural-rl_nas, liu2018darts-gradient}. However, traditional NAS often neglects hardware constraints, such as memory and compute efficiency, which are critical for IoT deployment.

HW-NAS extends NAS by incorporating hardware-specific metrics into optimization \cite{hw-nas_survey_benmeziane2021comprehensive}. It has been applied across domains to generate models optimized for constrained devices \cite{garavagno2024affordable, sinha2024multi_hw-nas, ragusa2024combining, ragusa2024compression}. These methods align network design with target platform requirements, enabling efficient, accurate models in real-world use. \added{The study in}~\cite{our_chehade2024tiny} explored HW-NAS for TC, and this paper expands on that by evaluating HW-NAS-optimized models on microcontrollers across multiple tasks.

\section{Methodology}

\subsection{Hardware-Constrained Optimization}

Deploying DNNs on microcontrollers demands a balance between performance and hardware limits. The HW-NAS framework optimizes architectures for session-level TC while ensuring compatibility with microcontroller-based platforms. The optimization problem seeks to maximize validation accuracy subject to typical embedded-system constraints:

\begin{maxi}|s|
    {a \in \mathcal{A}} {\text{Accuracy}_{\text{val}}(w^*(a), a)}
    {}{}
    \addConstraint{w^*(a) = \argmin_w \mathcal{L}_{\text{train}}(w, a)}
    \addConstraint{|P(a)| < F_{Th}}     
    \addConstraint{|T(a)| < R_{Th}}     
    \addConstraint{\text{Flops}(a) < \text{Flops}_{Th}}
    \label{eq:optprob}
\end{maxi}

where $\mathcal{A}$ represents the search space of candidate architectures\added{, and $a \in \mathcal{A}$ is a specific architecture}, and $w$ are the optimized model weights trained by minimizing the loss function $\mathcal{L}_{\text{train}}$. The hardware constraints include the parameter count \( |P(a)| \) (limited by Flash memory \( F_{Th} \)), maximum tensor size \( |T(a)| \) (RAM limit \( R_{Th} \)), and computational cost \( \text{Flops}(a) \) (bounded by \( \text{Flops}_{Th} \)). These thresholds are derived from prior works, such as \cite{ragusa2024combining, ragusa2024compression} to ensure efficient deployment on microcontrollers with strict memory and computation budgets. \added{Accordingly, the objective is solved via a constrained search that approximates a local minimum in the architecture space.
}

\subsection{1D-CNN Search Space Design}  

Prior work demonstrates that 1D-CNNs outperform 2D-CNNs for ETC tasks \cite{wang2017end}. Additionally, studies such as \cite{wang2017end, lotfollahi2020deep, wang2017malware} validate the effectiveness of 1D-CNNs in achieving strong accuracy for ETC. Moreover, 1D-CNNs are more efficient than RNNs or Transformers, especially when optimized for hardware-aware objectives. Motivated by these results, we design our search space around 1D-CNN architectures.

The search space comprises modular blocks, each containing a 1D convolutional layer, batch normalization, ReLU activation, pooling, and dropout. Hyperparameters such as kernel size, filter count, and stride \added{are automatically selected during the NAS process to control complexity and ensure efficient exploration.} \added{A full description of the search space setup used in our implementation is provided in Section~\ref{sec:experimental_setup}.}

\subsection{Evolutionary Architecture Search}  \label{algorithm}
The HW-NAS framework employs an evolutionary algorithm to explore the architecture space, guided by hardware constraints. This approach builds on prior work showing the success of evolutionary strategies for NAS \cite{1000papersNAS, ragusa2024combining}. Starting from an initial parent architecture \(a_0\), the algorithm generates candidates (children) via random mutations. Mutations include adding/removing blocks, adjusting hyperparameters (e.g., filters, kernel size), or modifying layer configurations. %
Each candidate is evaluated against hardware constraints; \added{invalid ones are discarded, and mutation continues until the required number of valid architectures is reached for training.} \added{Architectures exceeding a predefined depth limit are not generated.}

The process iterates over \(N_g\) generations, producing \(N_c\) candidates per generation. %
Valid candidates are trained on \(\mathcal{X_T}\) and evaluated on \(\mathcal{X_V}\). The highest-performing architecture becomes the parent for the next generation, repeating until convergence.  
The overall procedure is outlined in Algorithm~\ref{alg:proc}.  

\begin{algorithm}
    \caption{Evolutionary NAS with Hardware Constraints}
    \label{alg:proc}
    \textbf{Input:} Training set $\mathcal{X_T}$, validation set $\mathcal{X_V}$, search space $\mathcal{A}$, parent configuration $a_p$, child generator function $a_c=R_m(a_{p})$ number of children per generation $N_c$, total generations $N_g$ \\
    \textbf{Output:} Optimized architecture $a^*$  

    \begin{algorithmic}[1]
        \State \textbf{Initialize} parent architecture $a_p = a_0$
        \For{$g = 1$ to $N_g$}
            \For{$c = 1$ to $N_c$}
                \State \textbf{Mutation}: Generate child $a_c = R_m(a_p)$
                \If{$a_c$ satisfies hardware constraints}
                    \State \textbf{Training}: Train $a_c$ on $\mathcal{X_T}$
                \EndIf
            \EndFor
            \State \textbf{Selection}: Choose next parent $a_p = \argmax_{a_c} E(a_c, \mathcal{X_V})$
        \EndFor
        \State \textbf{Return} final architecture $a^* = a_p$
    \end{algorithmic}
\end{algorithm}

\section{Experimental Setup}

\subsection{Dataset and Traffic Classification Tasks}  
The ISCX VPN-nonVPN dataset \cite{iscxvpnnonvpn} is used in this study as a benchmark for TC. It consists of approximately 30GB of packet capture (pcap) files, covering various applications and categorized into VPN and Non-VPN traffic. Traffic is labeled according to application and usage type, supporting multiple classification tasks. The dataset has been widely adopted in DL-based TC research \cite{wang2017end, lotfollahi2020deep, lu2021iclstm, seydali2023cbs, zou2024novel_residual}.  

Several classification tasks are defined based on this dataset, as detailed in Table~\ref{tab:experiment_descriptions}. The table presents these tasks in rows, specifying their classification objectives and the number of output classes in the corresponding columns. The VPN-NonVPN task, which distinguishes VPN traffic from Non-VPN traffic, serves as the primary optimization task for HW-NAS due to its broad encapsulation coverage. The optimized model is subsequently evaluated across different classification tasks, including VPN detection, traffic type identification, and network usage categorization, to assess its adaptability.

\begin{table}[b]
\centering
\caption{Traffic Classification Tasks}
\label{tab:experiment_descriptions}
\renewcommand{\arraystretch}{1.05}
\small
\resizebox{\columnwidth}{!}{%
\begin{tabular}{|c|c|c|}
\hline
\textbf{Experiment} & \textbf{Description} & \textbf{Classes} \\
\hline
VPN-NonVPN & VPN and Non-VPN traffic classification & 11 \\
\hline
VPN-Diff & Protocol encapsulation detection & 2 \\
\hline
VPN-Type & VPN traffic type classification & 6 \\
\hline
NonVPN-Type & Non-VPN traffic type classification & 5 \\
\hline
Traffic-Cat & Network usage categorization & 6 \\
\hline
\end{tabular}%
}
\end{table}

\subsection{Preprocessing Pipeline}
A unified preprocessing pipeline is applied to ensure consistency across all classification tasks.
Raw traffic is segmented into sessions, where each session comprises a sequence of bidirectional packets exchanged between a source and destination IP address, with matching port numbers and protocols. Packet data is processed as raw bytes.

Data cleaning is applied to enhance model robustness. Data-link layer headers, including MAC addresses, are removed, and IP addresses are anonymized to prevent models from memorizing session-specific patterns. Non-payload packets, such as SYN, ACK, and FIN flags, as well as redundant DNS segments, are filtered out following prior research findings \cite{seydali2023cbs}.  

For input standardization, each session is normalized to a fixed length of 784 bytes following the approach of \cite{wang2017end, wang2017malware}. Sessions exceeding this length are truncated, while shorter ones are padded with null bytes to maintain consistency across inputs. The resulting byte sequences are then scaled to the range [0,1] to align with DL model expectations.

\subsection{HW-NAS Implementation and Training Configuration} \label{sec:experimental_setup}
 
The HW-NAS framework is executed on a workstation equipped with an Nvidia 2080 Ti GPU, leveraging Keras and TensorFlow for model training. It takes preprocessed session data as input and optimizes architectures based on accuracy while adhering to predefined hardware constraints on memory usage, FLOPs, and RAM. 

Therefore, since our goal is to deploy the model on microcontrollers, we first analyzed prior TC models, estimating their resource usage in Keras based on reported architectural details. This analysis revealed inefficiencies and guided our constraint selection. Given that the model is designed to run on both the STM32F746G-DISCO and the more constrained Nucleo-F401RE, we set thresholds to fit the latter’s stricter resource limitations (see Table~\ref{tab:hardware_specs} in Section~\ref{deployment}).

A holdout validation set comprising 20\% of the training data is used to guide the selection of the architecture. Training is performed for up to 100 epochs \added{using the Adam optimizer with} an initial learning rate of $10^{-3}$, a batch size of 128, and learning rate decay triggered when validation loss plateaus. Early stopping is applied to prevent overfitting, and each architecture is trained three times using a multi-start approach for improved stability.  

The search in HW-NAS spans 100 generations, with each generation producing 10 candidate architectures. The search space includes filters ranging from 16 to 140, kernel sizes between 3 and 7, and strides varying from 1 to 6. Dropout rates range from 0.1 to 0.5, with pooling operations selectable as max or average (pool size 2–3). Padding options include “same” \added{(output size preserved)} and “valid” \added{(no padding)}. The architecture that achieves the highest validation accuracy is selected as the final model.  

To evaluate generalization, the selected model is retrained in multiple TC tasks, with training extended to 200 epochs where necessary. Early stopping remains in effect, ensuring that training stops when no further improvements are observed.

\section{Results and Evaluation}

\subsection{HW-NAS-Optimized Architecture}  
The HW-NAS framework automatically designed a 1D-CNN architecture tailored for session-level TC. The final model features three convolutional layers with a decreasing number of filters: it starts with 129 filters (kernel size 7, stride 5), followed by 110 filters (kernel size 4, stride 2), and ends with 38 filters (kernel size 7, stride 2). The architecture incorporates average and max-pooling operations to adjust intermediate tensor dimensions. It finalizes with a global average pooling layer and a dense softmax classifier.

This design is highly efficient, featuring \textbf{88.26K parameters} (353KB Flash memory), a maximum tensor size of \textbf{20.12K} (80.5KB RAM), and \textbf{10.08M FLOPs}. These specifications fit both typical IoT constraints (e.g., \(\leq\)512\,KB Flash, \(\leq\)128\,KB RAM) and our target microcontrollers, enabling seamless edge deployment without GPU acceleration.

\subsection{VPN-NonVPN Classification and Hardware Efficiency}

Table~\ref{tab:sota_summary} compares the performance and hardware measures of the proposed model against state-of-the-art methods for the VPN-NonVPN task. The table rows list methods, categorized by their input type: session-level (Sess.), hybrid input methods (Hybrid), flow-based methods (Flows), and packet-level methods (Pack.). The columns display key metrics: classification accuracy (Acc.), F1 score (F1), parameter count (Par., in millions), maximum tensor size (M-Tens., in thousands), and floating-point operations (FLOPs, in millions).
The proposed model achieves 96.59\% accuracy and an F1 score of 96.54\%. This balance between performance and efficiency is notable compared to more resource-intensive models.

For instance, \cite{centime_maonan2021centime} achieves 99.79\% accuracy but requires 2.06M parameters and 3152.27M FLOPs—23 times more parameters and 312 times more FLOPs than the proposed model. Its tensor size is 200.70K, nearly 10 times larger, indicating higher memory use. Similarly, \cite{lu2021iclstm} achieves 98.10\% accuracy with 39.12M parameters, 251.06M FLOPs, and a tensor size of 151.51K, reflecting increases of 444 times in parameters, 25 times in FLOPs, and 7.5 times in tensor size.

Another session-based model \cite{wang2017end} is an early session-level classification approach, achieving 86.60\% accuracy with 5.83M parameters and 39.73M FLOPs. In contrast, it uses about 65 times more parameters and nearly 4 times more FLOPs but delivers much lower performance. This highlights the efficiency and accuracy gains of the proposed model.

Hybrid models \cite{seydali2023cbs} and \cite{cui2019session} achieve accuracies of 99.70\% and 99.10\%, respectively. \cite{seydali2023cbs} combines raw packet data with session-level statistics, while \cite{cui2019session} segments sessions using predefined thresholds. Both approaches increase computational complexity. \cite{seydali2023cbs} uses 138.45M parameters and 153.87M FLOPs, exceeding the proposed model’s hardware usage by factors of about 1570 and 15, respectively. Similarly, \cite{cui2019session} requires 6.80M parameters and 401.89M FLOPs, roughly 76 times more parameters and nearly 40 times the computational cost of our model.

Packet- and flow-based approaches, such as \cite{lotfollahi2020deep} and \cite{yao2019identification}, show varying hardware efficiencies but fall short of session-based model accuracy. For example, \cite{lotfollahi2020deep} achieves an F1 score of 93.00\% with 3.55M parameters and 208.19M FLOPs, indicating high computational costs without matching accuracy gains. Conversely, \cite{yao2019identification} requires less hardware in some setups but sacrifices accuracy, highlighting inherent trade-offs.

This comparison shows the proposed model achieves high accuracy while cutting computational demands, making it ideal for IoT environments with strict hardware constraints.

\enlargethispage{0.18in}

\begin{table}[http]
    \centering
    \caption{Performance and hardware comparison for VPN-NonVPN classification}
    \label{tab:sota_summary}
    \renewcommand{\arraystretch}{1.05}
    \large
    \resizebox{\columnwidth}{!}{%
    \begin{tabular}{|c|c|c|c|c|c|c|}
        \hline
        \makecell{\textbf{Meth.}} & \makecell{\textbf{Inp.}} & \makecell{\textbf{Acc.} \\ (\%)} & \makecell{\textbf{F1} \\ (\%)} & \makecell{\textbf{Par.} \\ (M)} & \makecell{\textbf{M-Tens.} \\ (K)} & \makecell{\textbf{FLOPs} \\ (M)} \\
        \hline
        \textbf{Proposal} & \textbf{Sess}. & \textbf{96.59} & \textbf{96.54} & \textbf{0.09} & \textbf{20.12} & \textbf{10.08} \\
        \cite{song2019encrypted} & Sess. & - & 91.80 & 0.22 & 313.60 & 267.22 \\
        \cite{zou2024novel_residual} & Sess. & 99.64 & 99.64 & 1.24 & 25.09 & 28.96 \\
        \cite{centime_maonan2021centime} & Sess. & 99.79 & 99.80 & 2.06 & 200.70 & 3152.27 \\
        \cite{wang2017end} & Sess. & 86.60 & - & 5.83 & 25.09 & 39.73 \\
        \cite{wang2020encrypted} & Sess. & 98.00 & 98.00 & 6.17 & 25.09 & 40.39 \\
        \cite{attention_spatio_hu2023network} & Sess. & - & 98.10 & 6.80 & 25.09 & 23.30 \\
        \cite{lu2021iclstm} & Sess. & 98.10 & 98.10 & 39.12 & 151.51 & 251.06 \\
        \hline
        \cite{seydali2023cbs} & Hybrid & 99.70 & 99.30 & 138.45 & 4500.00 & 153.87 \\
        \cite{cui2019session} & Hybrid & 99.10 & 99.30 & 6.80 & 102.40 & 401.89 \\
        \hline
        \cite{yao2019identification} & Flows & 91.20 & - & 1.35 & 2.00 & 0.86 \\
        \cite{yao2019identification} & Flows & 89.50 & - & 39.49 & 2.56 & 0.67 \\
        \hline
        \cite{lotfollahi2020deep} & Pack. & - & 93.00 & 3.55 & 100.00 & 208.19 \\
        \hline
    \end{tabular}%
    }
\end{table}

\subsection{Generalization Across Traffic Classification Tasks}

\begin{figure}[http]
    \centering
    \includegraphics[width=0.8\linewidth,height=7cm]{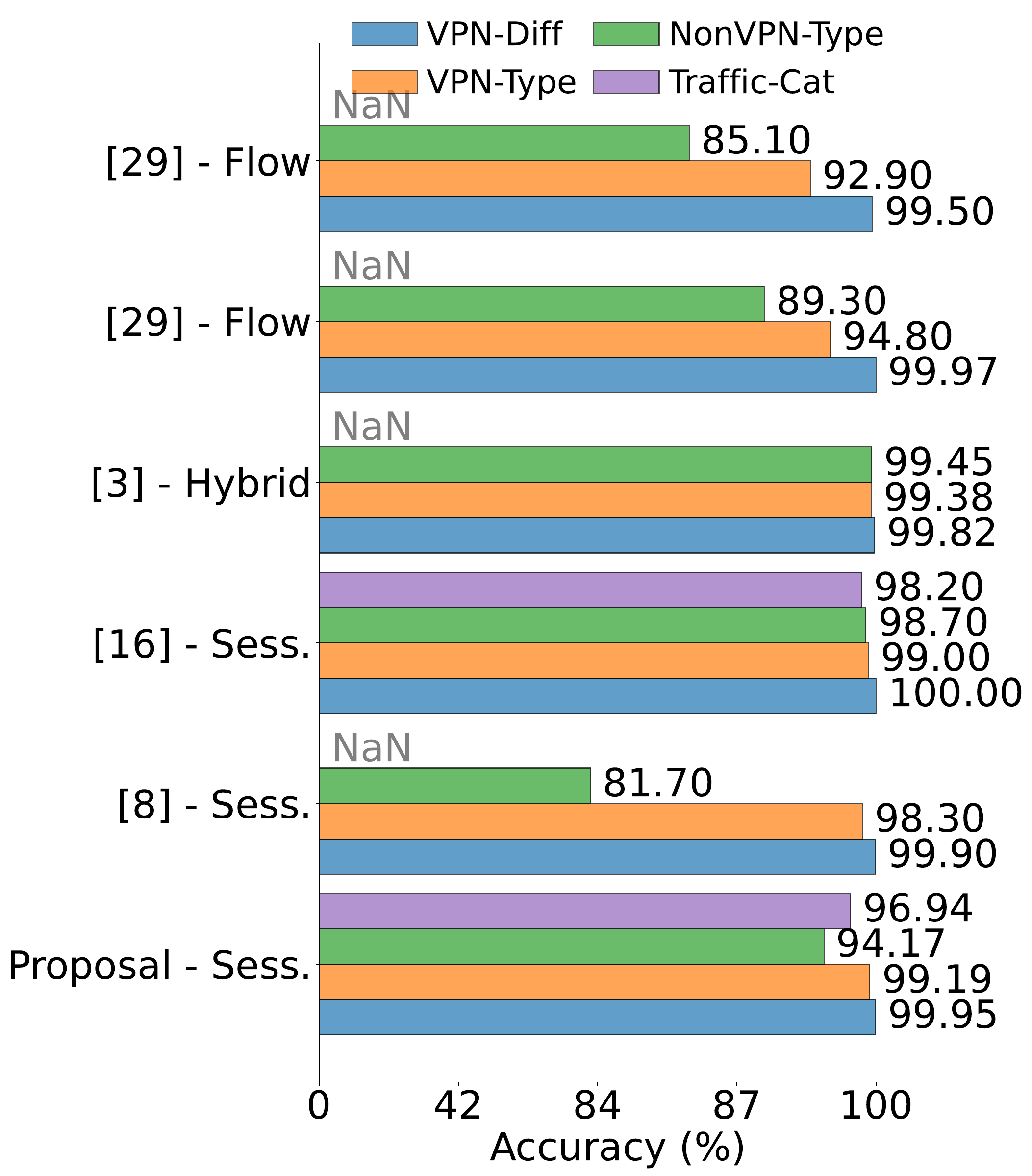}
    \caption{Accuracy across traffic classification tasks.}
    \label{fig:generalization_results}
\end{figure}

To evaluate the generalizability of our model, we tested it on four additional TC tasks beyond VPN-NonVPN: VPN-Diff, VPN-Type, NonVPN-Type, and Traffic-Cat (see Figure~\ref{fig:generalization_results}). 

The model achieves 99.95\% accuracy in VPN-Diff, matching \cite{lu2021iclstm} and outperforming \cite{seydali2023cbs} (99.82\%). It reaches 99.19\% in VPN-Type, surpassing \cite{yao2019identification} (94.80\%, 92.90\%) and \cite{wang2017end} (98.30\%). Similarly, for NonVPN-Type, it scores 94.17\%, outperforming \cite{yao2019identification} (89.30\%, 85.10\%) and \cite{wang2017end} (81.70\%). In Traffic-Cat, it attains 96.94\%, slightly below \cite{lu2021iclstm} (98.20\%) but with substantially lower hardware cost. These results confirm our architecture’s robustness and adaptability across diverse TC tasks, despite being optimized only for VPN-NonVPN.

\section{Deployment and Energy Efficiency Evaluation}

\subsection{Deployment Context} \label{deployment}

Efficient deployment on resource-constrained IoT devices is essential for practical session-level TC. To validate the HW-NAS optimized model's real-world applicability, we deployed it on two microcontroller boards, STM32F746G-DISCO and Nucleo-F401RE, which represent high-performance and cost-sensitive options, respectively. These boards embody typical embedded environments with strict memory, computational, and energy constraints\added{, and the model’s compact footprint supports broader deployment across similar low-power devices.}

Table~\ref{tab:hardware_specs} outlines the key specifications of both devices. The rows detail microcontroller architecture, memory capacity, and clock speed, while the columns differentiate the STM32F746G-DISCO and Nucleo-F401RE platforms. These specifications provide context for interpreting the performance and energy consumption results.

\begin{table}[b]
    \centering
    \caption{Device hardware specs for model deployment}
    \label{tab:hardware_specs}
    \renewcommand{\arraystretch}{1.05}
    \small
    \resizebox{\columnwidth}{!}{%
    \begin{tabular}{|c|c|c|}
        \hline
        \textbf{Specification} & \textbf{STM32F746G-DISCO} & \textbf{Nucleo-F401RE} \\
        \hline
        Microcontroller & Arm Cortex-M7 & Arm Cortex-M4 \\
        \hline
        Flash Memory & 1 MB & 512 KB \\
        \hline
        RAM & 340 KB & 96 KB \\
        \hline
        Clock Speed & 216 MHz & 84 MHz \\
        \hline
        Power Supply & 5V & 5V \\
        \hline

    \end{tabular}%
    }
\end{table}

\subsection{Model Deployment and Evaluation}

The optimized 1D-CNN model was quantized using 8-bit integer quantization and converted to TensorFlow Lite format. STM32Cube.AI was then used to translate the quantized model into C code for deployment on STM32 microcontrollers. Inference performance was evaluated by executing 1000 consecutive inferences on each device (with the reported inference times representing the average). Current and voltage were recorded using a USB power meter connected between the development board and the host PC during both idle and active inference phases. Power consumption during inference was calculated based on the additional current drawn relative to the idle state, and energy per inference was computed from this additional power and the inference latency.

Table~\ref{tab:quantization_results} compares the model's accuracy and F1 scores before and after quantization across TC tasks. The rows correspond to tasks, while the columns report accuracy and F1 score for the original Float32 model and the INT8 quantized version. The quantization process led to minor reductions in accuracy and F1 scores across most tasks. VPN-Diff and VPN-Type maintain nearly identical performance post-quantization, while other tasks, such as VPN-NonVPN and Traffic-Cat, exhibit drops between 1\% and 2\% in accuracy and F1 scores. This minimal degradation underscores the model's suitability for deployment on resource-constrained devices.

\begin{table}[b]
    \centering
    \caption{Accuracy and F1 before and after quantization}
    \label{tab:quantization_results}
    \renewcommand{\arraystretch}{1.05}
    \scriptsize
    \resizebox{\columnwidth}{!}{%
    \begin{tabular}{|c|c!{\vrule width 0.8pt}c|c!{\vrule width 0.8pt}c|}
        \hline
        \textbf{Task} 
        & \multicolumn{2}{c|}{\textbf{Float32}} 
        & \multicolumn{2}{c|}{\textbf{INT8 Quantized}} \\
        \cline{2-5}
         & Acc (\%) & F1 (\%) 
         & Acc (\%) & F1 (\%) \\
        \hline
        VPN-NonVPN & 96.59 & 96.54 & 94.73 & 94.11  \\
        \hline
        VPN-Diff & 99.95 & 99.95  &99.95  &99.13  \\
        \hline
        VPN-Type &99.19  & 99.19  &99.29  &97.43  \\
        \hline
        NonVPN-Type & 94.17 & 94.19 & 93.18 & 92.51  \\
        \hline
        Traffic-Cat & 96.94 & 96.98  &95.71  &94.83  \\
        \hline
    \end{tabular}%
    }
\end{table}

\subsection{Power and Energy Consumption}

Table~\ref{tab:energy_results} summarizes inference latency, power draw, and energy per inference for both devices. The STM32F746G-DISCO achieves an average inference latency of 31.43~ms and an energy consumption of 7.86~mJ per inference, while the Nucleo-F401RE exhibits a slower latency of 115.40~ms and higher energy consumption (29.10~mJ per inference). Despite similar power draws (0.25~W), the STM32F746G-DISCO's higher clock speed and larger memory yield more efficient processing. Still, the Nucleo-F401RE remains appealing for large-scale, cost-sensitive IoT deployments. \added{For session-level TC, such latency is acceptable given the aggregation delay already incurred during packet collection.
}

Overall, these results confirm that the HW-NAS optimized model meets the high-performance, low-energy demands of resource-constrained microcontrollers, balancing accuracy, latency, and cost. This validates its applicability in both performance-oriented and budget-limited scenarios.

\begin{table}[htbp]
    \centering
    \caption{Inference latency, power, and energy consumption}
    \label{tab:energy_results}
    \renewcommand{\arraystretch}{1.07}
    \footnotesize
    \resizebox{\columnwidth}{!}{%
    \begin{tabular}{|c|c|c|}
        \hline
        \textbf{Metric} & \textbf{STM32F746G-DISCO} & \textbf{Nucleo-F401RE} \\
        \hline
        Latency (ms) & 31.43 & 115.40 \\
        \hline
        Power (W) & 0.25 & 0.25 \\
        \hline
        Energy (mJ) & 7.86 & 29.10 \\
        \hline
    \end{tabular}%
    }
\end{table}

\section{Conclusion}
This study leveraged a HW-NAS approach to develop a lightweight 1D-CNN for session-level TC on resource-constrained microcontrollers. Our experiments confirm that the optimized model achieves high accuracy while reducing computational and memory overhead. It also generalizes well across diverse TC tasks, demonstrating broad applicability. Deployments on both a high-performance board (STM32F746G-DISCO) and a cost-sensitive alternative (Nucleo-F401RE) validate its real-world efficiency, with low inference latency and energy consumption across platforms. Future work will expand deployment to additional embedded systems and explore hardware-aware optimization for other cybersecurity applications, such as intrusion and anomaly detection, to advance real-time, decentralized network defense solutions.

\bibliography{references_adel}

\end{document}